\documentclass[conference]{IEEEtran}
\IEEEoverridecommandlockouts 
\usepackage{cite}
\usepackage{amsmath,amssymb,amsfonts}
\usepackage{algorithmic}
\usepackage{graphicx}
\usepackage{textcomp}
\usepackage{siunitx}
\usepackage{verbatim}
\usepackage[table,xcdraw]{xcolor}
\usepackage{comment}

\newcommand*\rot{\rotatebox{90}}
\newcommand*\OK{\checkmark}    
    
\begin{document}

\title{A Survey of Wearable Devices Pairing Based on Biometric Signals}



\author{
  \IEEEauthorblockN{Jafar Pourbemany, Ye Zhu}
\IEEEauthorblockA{\textit{Department of Electrical Engineering and Computer Science} \\
\textit{Cleveland State University}\\
Cleveland, USA \\
pourbemany@ieee.org, y.zhu61@csuohio.edu}
\and
\IEEEauthorblockN{Riccardo Bettati}
\IEEEauthorblockA{\textit{Department of Computer Science and Engineering} \\
\textit{Texas A\&M University}\\
Texas, USA \\
bettati@tamu.edu}
}

\maketitle
\begin{abstract}
With the growth of wearable devices, which are usually constrained in computational power and user interface, this pairing has to be autonomous. Considering devices that do not have prior information about each other, a secure communication should be established by generating a shared secret key derived from a common context between the devices. Context-based pairing solutions increase the usability of wearable device pairing by eliminating any human involvement in the pairing process. This is possible by utilizing onboard sensors (with the same sensing modalities) to capture a common physical context (e.g., body motion, gait, heartbeat, respiration, and EMG signal). A wide range of approaches has been proposed to address autonomous pairing in wearable devices. This paper surveys context-based pairing in wearable devices by focusing on the signals and sensors exploited. We review the steps needed for generating a common key and provide a survey of existing techniques utilized in each step. 

\end{abstract}



\section{Introduction}
\label{sec:introduction}
\begin{figure*}[t!]
    \centering
    \includegraphics[width=9cm]{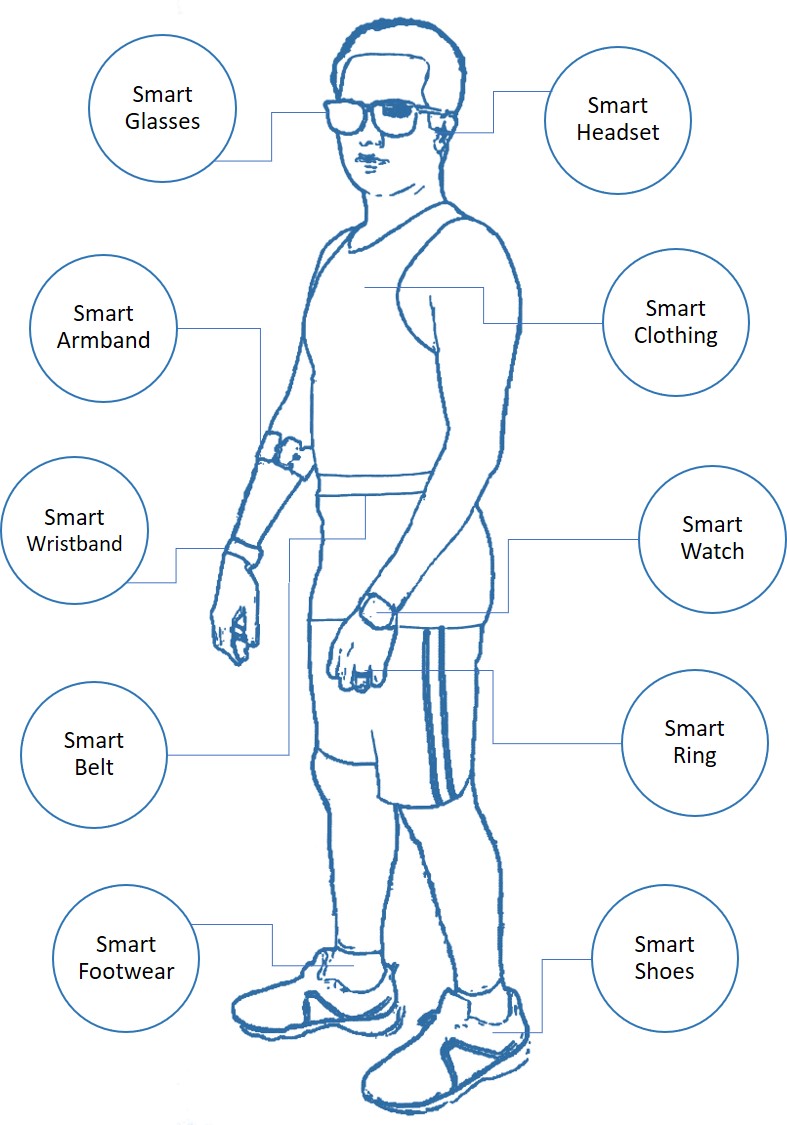}
    \caption{Different types of wearable devices.}\label{wearable_devices}
 \end{figure*}

Smart wearable devices can collect a variety of information about human activities and behaviors which makes them popular in clinical medicine and health care, health management, workplace, education, and scientific research. The wearable market is diversified with hundreds of products, including smartwatches, smart wristbands, smart glasses, smart jewelry, smart straps, smart clothes, smart belts, smart shoes, smart gloves, skin patches, and even implanted medical devices (IMD) \cite{Guk2019, Vhaduri2019, chong2014survey, Seneviratne2017, Bianchi2016, JohnDian2020, Poongodi2020, Pantelopoulos2010}. As Fig. \ref{wearable_devices} illustrates, there are many diverse types of wearable devices for different body parts that collect various physiological data and possibly make intelligent decisions based on the collected physiological data.
Wearable devices are exploited in a wide range of applications \cite{JohnDian2020}. For example, wearable devices can be used as health monitoring systems and health treatment systems. They can monitor vital signs like heart rate \cite{jayanth2017wearable,majumder2019energy, brezulianu2019iot}, respiratory rate \cite{milici2016wireless, shah2019cloud, mahbub2017low, naranjo2018smart}, body temperature \cite{wan2018wearable, yoshida2018development}. Others collect parameters like blood pressure \cite{lamonaca2019overview}, blood oxygen \cite{murali2018pulse}, blood glucose \cite{sargunam2019iot}, to detect disorders. Some wearable devices also can help disabled patients to recover certain physical functions \cite{nave2018smart, yang2018iot, agyeman2019design}.
Wearable devices are applied beyond healthcare. For example, wearable devices can be used to recognize daily physical activities \cite{qi2018hybrid, castro2017wearable, huang2018tribomotion, lu2018wearable, atallah2011sensor, mathie2004classification, mannini2010machine} or the activities that are related to specific sports \cite{mencarini2019designing, wang2016badminton, raina2017combat, maglott2017differences, bogers2017design, wang2015stance, wang2016canoesense, pansiot2010swimming, bachlin2009swimmaster, hakkila2016mydata, haladjian2017vihapp, niforatos2017augmenting, chi2004killer, pan2018designing, havlucu2017understanding, lapinski2009distributed, ghasemzadeh2009sport}. Other applications areas for wearable devices include payment management \cite{seneviratne2017survey, wang2019wristspy}, unlocking vehicles \cite{nguyen2017smartwatches}, keeping sensitive information (like passwords) \cite{nguyen2017smartwatches}, controlling paired devices \cite{kumar2016authenticating}, subject tracking \cite{shit2018location, kim2012smartphone}, fall detection \cite{saadeh2019patient, saadeh2017wearable, ren2019research, otanasap2016pre, nyan2008wearable}, drowsiness detection \cite{choudhary2016survey, chang2018design, dhole2019novel}, environment monitoring \cite{wu2018self, serbanescu2017smart}, virtual and augmented reality \cite{perez2016optimization, maisto2017evaluation}, and spying \cite{wang2019wristspy}. 

Based on their application, the wearable devices collect, analyze, and store the data. In some cases, the wearable device (e.g., pacemaker) may be equipped with actuators which it may use to control important physiological functions of the wearer. Often, they need to share data or commands with the base station or other wearable devices. Data/command shared between wearable devices are usually sensitive like stored credentials of the user or commands to adjust an IMD. Therefore, the communication should be protected by encryption and the devices need to have a common key for encryption to ensure that the process has not been compromised by an attacker.

Many wearable devices use Bluetooth or Wi-Fi to connect to either other smartphones or other wearables, and the Internet.
To be able to send and receive data the new wearable device must establish a connection with the other device; this process is called pairing (also known as binding, coupling, bonding, or association \cite{chong2014survey}). For example, the user initially needs to pair a new smartwatch with their smartphone over Bluetooth or NFC channel before use. Some wearable medical devices such as IMDs need to be paired to their external controllers to receive updates.

Traditional pairing techniques often need user actions can have various forms. A common approach is used in Bluetooth pairing: The user selects a target device from a list of available devices and then possibly uses a PIN for additional authentication \cite{Fomichev2017}. However, the approaches that require an initial stage of network setup are not scalable as the number of wearables increases \cite{unar2014review}. PIN-based approaches need interaction with the display, which may either be inconvenient or even impossible in many wearables \cite{zeng2017wearia, unar2014review}. Furthermore, PIN-based pairing is vulnerable to observation attacks such as shoulder surfing \cite{unar2014review}. Public key cryptography (PKC), on the other hand, cannot be used to create a secure key on wearable devices because it requires a public key infrastructure (PKI) \cite{shim2015survey}. In addition, it requires expensive computing methods that are not suitable for resource-limited IoT devices. To mitigate these limitations, various techniques have been proposed that take advantage of common features in different wearable devices. 
Pairing based on biometric signals is a natural choice because wearable devices are attached to the same body. Both behavioral biometrics (e.g., step counts) and physiological biometrics (e.g., heart rate) are used for wearable device pairing and authentication. 

In this article, we survey context-based techniques that can be used to achieve automatic wearable pairing based on them. In Section II, we review a selection of survey papers on biometric-based approaches. Section III discusses the signals used for pairing smart wearables. In Section IV, different steps of biometric-based pairing mechanism are surveyed. In Section V, we investgate the most common adversary models. Limitations and challenges are presented in Section VI. Finally, Section VII concludes the paper.


\section{Related studies on biometric-based approaches}
Biometric signals are widely used for device authentication. Variety of context-based approaches are used in device authentication to verify user identity \cite{sarier2009survey} on a unique device.

There are several surveys that have investigated different aspects of the security and authentication in the wireless body area network (WBAN) \cite{cavallari2014survey,movassaghi2014wireless}. In Table \ref{related_surveys}, we summarize the topics raised in surveys over the last ten years. Security essentials and existing attacks against WBAN have been addressed in Javadi, and Razzaque \cite{Javadi2013}. They also discussed the significant constraints and challenges of security mechanisms. A detailed review of authentication schemes is conducted by Masdari and Ahmadzadeh \cite{masdari2016}. The authors provided a taxonomy for authentication modalities that classifies it into three categories: biometric-based, channel-based, and cryptography-based. Mainanwal \emph{et al.} \cite{Mainanwal2015} summarized the pros and cons of several security and privacy techniques and discussed the threats and challenges. Sawaneh \emph{et al.} \cite{Naik2017} surveyed the views on security and privacy essentials for WBAN. Keystroke dynamics are addressed in \cite{Bhatt2013,Karnan2011}. Abuhamad \emph{et al.} \cite{Abuhamad2021} investigated the authentication protocols and OS-related security of smartphones’ users using behavioral biometrics. 

The survey by Janabi \emph{et al.} \cite{Al-Janabi2017} reviewed the major security and privacy problems in WBAN for healthcare applications, along with their state-of-the-art security solutions. The paper provides significant highlights on open issues and future research directions in WBANs. Aman and Shah \cite{Zou2017} explored the security problems of ubiquitous healthcare (U-Healthcare) related work. The survey by Narwal and Mohapatra \cite{Narwal2018} focuses on the analysis of authentication schemes in terms of main outcomes, strengths, and limitations. In addition to this, the authors discuss the architecture of WBAN, security essentials, and security attacks are discussed in detail. A taxonomy is proposed by Usman \emph{et al.} \cite{usman2018} that classifys entities involved in healthcare systems. Security challenges at all WBAN tiers have been studied. The authors also identified open issues and highlighted future research directions. In another work, the cryptographic solutions have been reviewed by Malik \emph{et al.} \cite{Malik2018}. They provided a general survey on major security essentials and conceivable assaults at different layers. 

Considering security view of the complete WBAN system, Morales \emph{et al.} \cite{morales2019} focused on various protocols in WBAN architecture and provided a detailed review of security requirements such as confidentiality, integrity, privacy, authentication, and authorization. Komapara and Holbl \cite{Kompara2018} reviewed security and key agreement of Intra-BAN communication. They categorised the existing key agreement schemes into traditional, physiological value-based, secret key-based, and hybrid key-based schemes. Furthermore, the authors provided the description of each class and analysed the security strength of BAN towards attacks. Nidhya and Karthik \cite{nidhya2019} emphasized the security attacks and security models at data collection, transmission, and data storage level. Also, they assessed the privacy requirements and reliability of healthcare systems. Joshi and Mohapatra \cite{Joshi2019} investigated design, functionalities, and workflow of the existing authentication schemes. They described the structure of authentication schemes and provided a detailed view on communication standard and design issues in WBAN. The paper also suggests methods to safeguard the key during key management.

Surveying security and Privacy issues in WBAN, Chaudhary \emph{et al.} \cite{Chaudhary2019} classified authentication schemes into four categories: Physiological Value-based, Channel-based, Proximity-based, and Cryptographic-based. Furthermore, they summarized various schemes from different categories in a tabular form to highlight the features of each scheme effectively. The survey by Hussain \emph{et al.} \cite{Hussain2019} conducted to provide greater insight into authentication schemes. All in all, the survey presented a detailed discussion on security features, security attacks, strengths, limitations, and performance of the authentication schemes. Roy \emph{et al.} \cite{roy2020} reviewed the major security and privacy issues in WSNs and WBANs. They conducted a comparative analysis of both the networks based on their features, architecture, applications and threats. In another work, Narwal et al \cite{Narwal2021} explored the authentication schemes in different categories. 

All the aforementioned studies surveyed a variety of methods used sensory signals to assist IoT device authentication. However, sensors data, especially biometric signals, can be used for pairing the wearable devices as well. Indeed, biometric signals are the common point between wearable device authentication and pairing, whereas the goals and applications are different. This survey focuses on sensor-based pairing in wearable devices to provide more details about the challenges and limitations of sensors in such devices.

\begin{table*}[]
\caption{Survey papers on the WBAN authentication - Content comparison}
\centering
\resizebox{\textwidth}{!}{%
\begin{tabular}{|l|l|l|l|l|l|l|l|l|l|l|l|l|}
\hline
Scheme & Year & \begin{tabular}[c]{@{}l@{}}Methods\\ classifications\end{tabular} & \begin{tabular}[c]{@{}l@{}}Knowledge-based\\    methods\end{tabular} & Motion & Gait & ECG & PPG & \begin{tabular}[c]{@{}l@{}}Advantages and\\  disadvantages\end{tabular} & \begin{tabular}[c]{@{}l@{}}Attack \\ senarios\end{tabular} & \begin{tabular}[c]{@{}l@{}}Future\\  works\end{tabular} \\ \hline
\cite{Karnan2011} & 2011 & $\blacksquare$ & $\blacksquare$ &  &  &  &  &  &  &  \\ \hline
\cite{Javadi2013} & 2013 &  & $\blacksquare$ &  &  &  &  &  & $\blacksquare$ &  \\ \hline
\cite{Bhatt2013} & 2013 & $\blacksquare$ & $\blacksquare$ &  &  &  &  &  &  &  \\ \hline
\cite{Mainanwal2015} & 2015 &  & $\blacksquare$ &  &  &  &  &  &  &  \\ \hline
\cite{masdari2016} & 2016 & $\blacksquare$ & $\blacksquare$ &  &  & $\blacksquare$ &  & $\blacksquare$ & $\blacksquare$ &  \\ \hline
\cite{Naik2017} & 2016 &  & $\blacksquare$ &  &  &  &  &  &  &  \\ \hline
\cite{AlAbdulwahid2016} & 2016 & $\blacksquare$ & $\blacksquare$ &  &  &  &  & $\blacksquare$ &  &  \\ \hline
\cite{Spolaor2016} & 2016 &  & $\blacksquare$ &  &  &  &  &  &  &  \\ \hline
\cite{Neal2016} & 2016 & $\blacksquare$ & $\blacksquare$ & $\blacksquare$ & $\blacksquare$ &  &  & $\blacksquare$ &  &  \\ \hline
\cite{Zou2017} & 2017 & $\blacksquare$ & $\blacksquare$ &  &  & $\blacksquare$ &  &  &  & $\blacksquare$ \\ \hline
\cite{Al-Janabi2017} & 2017 &  & $\blacksquare$ &  &  &  &  &  & $\blacksquare$ &  \\ \hline
\cite{Narwal2018} & 2018 &  & $\blacksquare$ &  &  & $\blacksquare$ &  & $\blacksquare$ & $\blacksquare$ &  \\ \hline
\cite{usman2018} & 2018 & $\blacksquare$ &  &  &  &  &  &  & $\blacksquare$ & $\blacksquare$ \\ \hline
\cite{Malik2018} & 2018 &  & $\blacksquare$ &  &  & $\blacksquare$ &  &  & $\blacksquare$ &  \\ \hline
\cite{morales2019} & 2018 &  & $\blacksquare$ &  &  &  &  &  & $\blacksquare$ &  \\ \hline
\cite{Kompara2018} & 2018 & $\blacksquare$ & $\blacksquare$ & $\blacksquare$ &  & $\blacksquare$ & $\blacksquare$ &  & $\blacksquare$ &  \\ \hline
\cite{nidhya2019} & 2019 &  & $\blacksquare$ &  &  & $\blacksquare$ &  &  & $\blacksquare$ &  \\ \hline
\cite{Joshi2019} & 2019 & $\blacksquare$ & $\blacksquare$ &  &  &  &  &  & $\blacksquare$ &  \\ \hline
\cite{Chaudhary2019} & 2019 & $\blacksquare$ & $\blacksquare$ &  &  & $\blacksquare$ &  &  &  &  \\ \hline
\cite{Hussain2019} & 2019 & $\blacksquare$ & $\blacksquare$ &  &  & $\blacksquare$ & $\blacksquare$ & $\blacksquare$ & $\blacksquare$ &  \\ \hline
\cite{roy2020} & 2020 &  & $\blacksquare$ &  &  &  &  &  & $\blacksquare$ &  \\ \hline
\cite{Narwal2021} & 2021 & $\blacksquare$ & $\blacksquare$ &  &  & $\blacksquare$ &  & $\blacksquare$ & $\blacksquare$ & $\blacksquare$ \\ \hline
\cite{Abuhamad2021} & 2021 & $\blacksquare$ & $\blacksquare$ & $\blacksquare$ & $\blacksquare$ &  &  & $\blacksquare$ &  &  \\ \hline
\end{tabular}%
}
\label{related_surveys}
\end{table*}

\section{Signals used for pairing}
\begin{figure*}[t!]
   \centering
   \includegraphics[width=15cm,height=12cm]{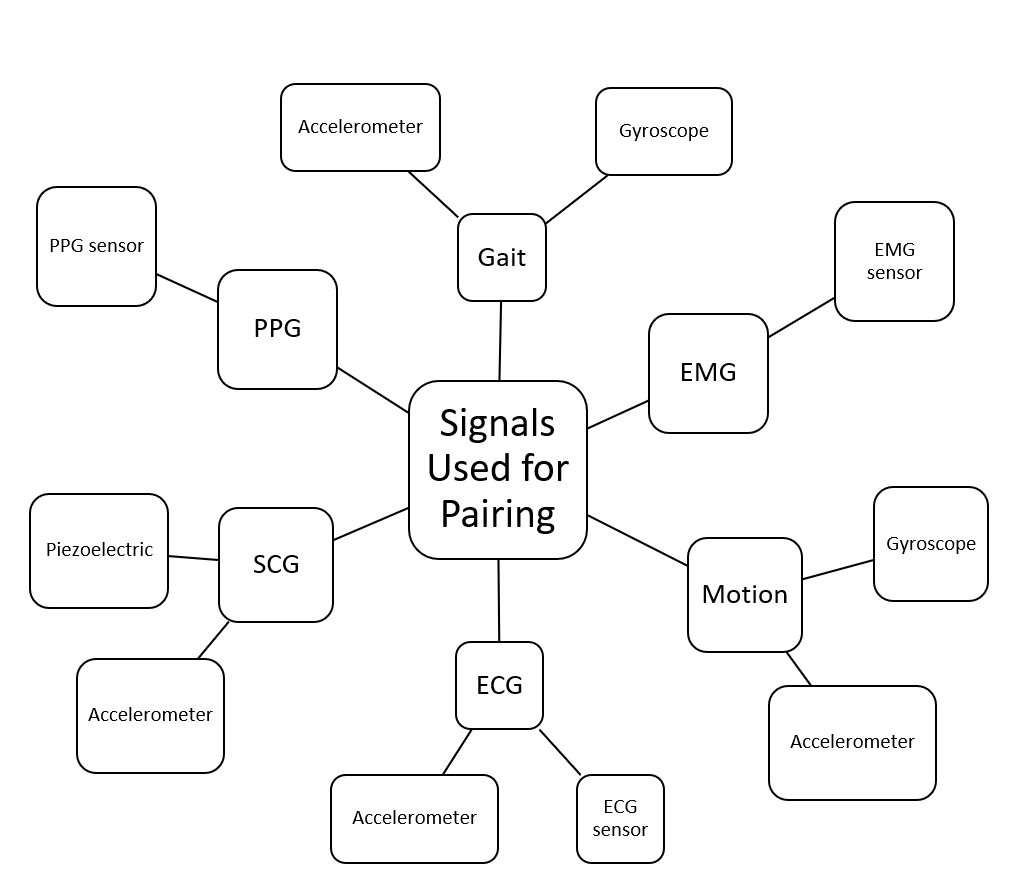}
   \caption{Signals used for pairing.}\label{signals}
\end{figure*}

Various types of sensors collect a variety of information from the human body and the environment. As Fig. \ref{signals} shows, several signals are used for context-base pairing, including motion, gait, ECG, PPG, EMG, and SCG. Indeed, these sensors can provide auxiliary out-of-band (OOB) channels \cite{mayrhofer2012uacap} as a feasible option to facilitate device pairing. We categorize and describe the signals used for wearable device pairing in recent papers in this section.


\subsection{Motion and position}
Magneto-Inertial Measurement Unit sensors (MIMU), including accelerometer, gyroscope, and magnetometer, are the most common sensors in today's wearable devices. These sensors detect the user's motion and heading of the device with respect to the Earth’s magnetic north pole. The accelerometer measures acceleration in three orthogonal spatial dimensions, x, y, and z, where each axis denotes either the vertical, forward-to-backward, or left-to-right dimensions in meter per second squared \cite{ferrero2015gait}. The gyroscope measures the angular rotation about each of these axes in radians per second \cite{fantana2015movement}. The magnetometer measures the strength of the local magnetic field along three orthogonal axes \cite{jin2015magpairing}. User's body movement can be modeled by using the information provided by these sensors. Hence, a variety of methods have been proposed to use such sensory data for authentication and pairing purposes.

For instance, in \cite{LaMarca2007a, Bichler2007a}, the authors proposed to shake the two mobile devices held in one hand; in this way, the accelerometer reading could be used to generate matched keys on either frequency domain \cite{LaMarca2007a} or time-domain \cite{Bichler2007a}. The authors demonstrate that the simultaneous shaking motion of two devices generates unique accelerometer readings that an adversary cannot easily mimic at a close distance. Shen \emph{et al.} \cite{Shen2018} proposed a method in which using a similar motion pattern of handshaking, two devices on the different bodies can be paired. Similarly, in \cite{jiang2019shake}, a handshake-based pairing scheme between wrist-worn smart devices are developed based on the observation that, by shaking hands, both wrist-worn smart devices conduct similar movement patterns. Hash functions and heuristic search trees were leveraged in \cite{Groza2012} to propose a key exchange protocol based on accelerometer data while the user shakes devices together. In another work, Yuzuguzel \emph{et al.} \cite{Yuzuguzel2015a} propose to derive a shared key based on finding a small set of effective features from shaking of two devices that are held together in one hand. Gorza \emph{et al.} \cite{Groza2020a} investigated the pairing of mobile devices based on shared accelerometer data under various transportation environments. Using multiple sensors (accelerometer and microphone), Cao \emph{et al.} \cite{Cao2019} presented a device-to-Device (D2D) communication that the user needs to hold two devices in one hand and randomly shake them for a few seconds. Jin \emph{et al.} \cite{jin2015magpairing} proposed a scheme to pair smartphones in close distance by exploiting correlated magnetometer readings. 

Table \ref{tbl_motion} shows a summary of features of several papers in the field of motion and position based device pairing.

\begin{table*}[]
\centering
\caption{Motion-based pairing papers and their features}
\begin{tabular}{|l|l|p{1cm}|p{0.7cm}|p{0.5cm}|p{0.7cm}|p{0.5cm}|p{0.4cm}|p{0.4cm}|p{0.5cm}|p{0.4cm}|p{0.4cm}|p{0.4cm}|p{1.5cm}|p{3cm}|}
\hline
\rot{\textbf{Scheme}} & \rot{\textbf{Year}} & \rot{\textbf{Sensor and Tools}} & \rot{\textbf{Key Length (bit)}} & \rot{\textbf{Duration (sec.)}} & \rot{\textbf{Success Rate}} & \rot{\textbf{Equal Error Rate   (EER)}} & \rot{\textbf{FAR}} & \rot{\textbf{FRR}} & \rot{\textbf{Bit Agreement}} & \rot{\textbf{Computation Time}} & \rot{\textbf{Energy   Consumption}} & \rot{\textbf{Frequency}} & \rot{\textbf{Subjects}} & \rot{\textbf{Device}} \\ \hline
\cite{Shen2018} & 2018 & Acc & 128 & 1 & \textgreater{}99\% & 1.60\% & \OK & \OK &  & \OK & \OK &  &  & Wrist worn smart   wearables \\ \hline
\cite{Bejder2020} & 2020 & Acc & 128 & 2.33 &  &  &  &  &  & \OK & \OK & 25 &  & LaunchPad CC1350 \\ \hline
\cite{jiang2019shake} & 2019 & Acc & 128 &  & close to 100\% & 1.5\% & \OK & \OK &  & \OK &  & 100 & 16 (8m 8f) & iPhone6 and iPhone8 \\ \hline
\cite{Groza2020a} & 2020 & Acc & 448 & 90 &  &  &  &  &  & \OK &  & 5 &  & LG Optimus L7 P700,   Samsung J5 \\ \hline
\cite{Cao2019} & 2019 & Acc, Mic, Speaker &  &  &  &  & \textless{}5\% & \textless{}2\% & \textgreater{}90\% & \OK & \OK & 100 & 10 (6m 4f) & HUAWEI MATE8 \\ \hline
\cite{Yuzuguzel2015a} & 2015 & Acc & 40 & 5 & 76\% & 4\% & \OK & \OK &  &  &  & 100 & 10 (5m 5f) &  \\ \hline
\cite{sen2020vibering} & 2020 & Acc, Vibration & 12.5 & 1 & 85.90\% & 5.00\% &  &  &  &  &  & 500 & 12 (5m 7f) & Arduino,   accelerometer MPU-6050 \\ \hline
\cite{Groza2012} & 2012 & Acc & 64 &  & 75\% &  &  &  &  &  &  & 80-90 &  & WiiMotes \\ \hline
\cite{LaMarca2007a} & 2007 & Acc & 128 & 3 &  &  & \OK & \OK &  &  &  & 600 & 51 (32m 19f) & ADXL202JE   accelerometers \\ \hline
\cite{Bichler2007a} & 2007 & Acc & 140 &  & 80\% &  &  &  &  &  &  & 200 & 10 &  \\ \hline
\cite{jin2015magpairing} & 2016 & Magnet- ometer & 128 & 4.5 & \textgreater{}= 90\% &  & \OK &  &  &  &  & 50 &  & Various kinds of   smartphones \\ \hline
\end{tabular}
\label{tbl_motion}
\end{table*}

\subsection{gait}
Gait recognition is the process of identifying an individual via how he or she walks based on wearable sensors data, especially motion sensors (e.g., accelerometer and gyroscope) \cite{derawi2010unobtrusive}. Due to the different properties of an individual’s muscular-skeletal structure, gait patterns are fairly unique among individuals \cite{zhong2014sensor}. Hence, it can be determined if two devices are carried by the same person \cite{Lester2004}. 
  
Various techniques exploit different features of gait to generate a common key for pairing wearable devices. Sun \emph{et al.} \cite{Sun2017} proposed a method to generate a symmetric key based on the timing information of gait. The authors used the IPI of consecutive gait as a common feature between the two devices. Schürmann \emph{et al.} \cite{Schurmann2018} presented a secure spontaneous authentication scheme that exploits correlation in acceleration sequences from devices worn or carried together by the same person to extract always-fresh secure secrets. In their method, BANDANA, they utilized instantaneous variations in gait sequences with respect to the mean. Walkie-Talkie \cite{xu2016walkie} is another shared secret key generation scheme that allows two legitimate devices to establish a common cryptographic key by exploiting users’ walking characteristics (gait). The authors exploit independent component analysis (ICA) for blind source separation (BSS) to separate accelerometer signals from different body movements such as arm swing and walk. In Gait-Key \cite{xu2017gait} Xu \emph{et al.} extended their method in Walkie-Talkie to examine the effect of multi-level quantization on the pairing success rate. In \cite{xu2019gait} the same authors also proposed using spatial alignment instead of using BSS. A usability analysis of four gait-based device pairing schemes \cite{Schurmann2018,xu2016walkie,Sun2017,Groza2012} are presented in \cite{Brusch2020}.
A summary of features of several gait-based pairing papers is shown in Table \ref{tbl_gait}.

\begin{table*}[]
\centering
\caption{Gait-based pairing papers and their features}
\begin{tabular}{|l|l|l|l|l|l|l|l|l|l|l|l|}
\hline
\rot{\textbf{Scheme}} & \rot{\textbf{Year}} & \rot{\textbf{Sensor and tools}} & \rot{\textbf{Key length (bit)}} & \rot{\textbf{Duration (sec.)}} & \rot{\textbf{Success rate}} & \rot{\textbf{Bit agreement}} & \rot{\textbf{Computation Time}} & \rot{\textbf{Energy   Consumption}} & \rot{\textbf{Frequency}} & \rot{\textbf{Subjects}} & \rot{\textbf{Device}} \\ \hline
\cite{Sun2017} & 2017 & Acc &  &  & 79\% & \OK &  &  & 100 & 5 & iPhone \\ \hline
\cite{Schurmann2018} & 2018 & Acc & 16 & 12 & \textgreater{}=75\% &  &  &  & 50 & 482 and 15 &  \\ \hline
\cite{xu2016walkie} & 2016 & Acc & 128 & 5 &  & \OK & \OK & \OK & 100 & 20 & Motorola E2 \\ \hline
\cite{xu2017gait} & 2017 & Acc & 128 & 4.6 & 98.30\% & \OK & \OK & \OK & 100 & 20 (14m 6f) & Motorola E3 \\ \hline
\cite{xu2019gait} & 2019 & Acc & 128 & 5 &  & \OK & \OK & \OK & 100 & 21 (14m 6f) & Motorola E4 \\ \hline
\end{tabular}
\label{tbl_gait}
\end{table*}

\subsection{ECG and PPG}
The heart-beat is a promising option for wireless body area networks (WBANs) authentication and key generating schemes because its properties are unique, and their features differ from person to person \cite{Thotahewa2014}. Heart-beat signals can be easily collected, and they are hard to copy by other people in comparison to simple pin codes. It is more secure than traditional methods because it requires a user to be available at the time of authentication and pairing process \cite{Sujatha2013, Wang2011}. Heart-beat signals can usually be collected by ECG and PPG sensors. ECG sensors collect the electrical activity of heart muscles through electrodes attached to the body. PPG sensors which can be attached to different parts of the body like the ear and finger, detect the blood level transforms in the microvascular cot of tissue \cite{Rundo2018}. It illuminates the body and measures transforms in light absorption as blood circulates in the body. The heart-beat signal can also be measured by seismocardiogram (SCG), which is the chest movement in response to the heart-beat. Accelerometers and piezo vibration sensors in the wearable devices can measure SCG as well \cite{Lin2019,wang2018unlock,ramos2012heart}. 
  
Various features extracted from heart-beat signals can be used for authentication and key generating purpose. The most important feature used in WBANs is heart rate variability (HRV) or R-R interval or inter-beat interval (IBI) or Inter-pulse Interval (IPI) \cite{karaa2015biomedical, Sufi2010, Okoh2015} indicates the time interval between consecutive heart-beats \cite{McCraty2015}. Indeed, the fluctuations of heart rate around an average rate are shown by HRV \cite{karaa2015biomedical}. As has been proven by several studies \cite{cherukuri2003biosec,Rostami2013,obrist2012cardiovascular}, HRV is highly random and can be used as a random source to generate keys. Since HRV is a unique characteristic for each person, it can be used as an authentication method to pair devices on the same body.

Rostami \emph{et al.} \cite{Rostami2013} proposed an HRV-based pairing method to authenticate external medical device controllers and programmers to IMDs.
The authors introduce a touch-to-access policy using a time-varying physiological value (PV) by ECG readings. They utilized statistical characterization of ECG for pairing wearable devices. Another pairing system called H2B is presented by Lin \emph{et al.} \cite{Lin2019}, which utilizes piezo sensors to detect heart-beat signals and generate a secret key.

\subsection{EMG}
The EMG or electromyogram signals are the electrical signal generated by contractions of human muscles. According to medical research \cite{merletti2004electromyography,devasahayam2012signals}, the EMG signal is a quasi-random process, i.e., the average value of EMG is correlated to the generated force of the muscle, but it has a random amplitude variation under a given force. In other words, there are stochastic variations of EMG amplitude for a unique gesture and force. Therefore, the EMG signals can be used as a secure source to generate secret keys in physically close contact for some wearable devices like Myo armband \cite{Myo}, Athos gear \cite{Athos}, and Leo smart band \cite{Leo}. Since detecting this kind of signal needs physical contact in proximity, it is extremely difficult for an adversary to perform an eavesdropping attack. EMG-KEY is an EMG-based method proposed by Yang \emph{et al.} \cite{Yang2016} which leverages EMG variation signal to generate a secret key for pairing two wearable devices.

A summary of different signals and sensors used for wearable device pairing is shown in Table \ref{tbl_signals}.

\begin{table*}[]
\centering
\caption{Different types of signals and sensors used for pairing}
\begin{tabular}{|l|p{1.2cm}|l|l|p{1cm}|p{5cm}|p{3cm}|p{2.3cm}|}
\hline
\textbf{Scheme} & \textbf{Name} & \textbf{Year} & \textbf{Signal} & \textbf{Sensor and tools} & \textbf{Description} & \textbf{Technique} & \textbf{Attack senarios} \\ \hline
\cite{Bejder2020} & SHAKE & 2020 & Motion & Acc & Holding   devices together and shaking them. SHAKE leverages the shared acceleration   patterns to generate the same key in both devices. & Shaking   together & Eavesdroping,  mimic attack \\ \hline
\cite{Groza2020a} & Multi-Modal   Transport & 2020 & Motion & Acc & Pairing   of mobile devices based on accelerometer data under various transportation   environments & Detecting   the same variation in different devices & Full   control of the communication channel \\ \hline
\cite{sen2020vibering} & VibeRing & 2020 & Motion & Acc,   vibration & Use   of vibration, generated by a custom Ring, as an out-of-band communication   channel to unobtrusively share a secret with a Thing & Detecting   vibration of ring & Impersonating,  eavesdropping \\ \hline
\cite{jiang2019shake} & Shake   to Communicate & 2019 & Motion & Acc & A secure   handshake acceleration-based pairing mechanism for wrist worn devices & Handshaking & Passive   attack,  active attacks \\ \hline
\cite{Cao2019} & Sec-D2D & 2019 & Motion & Acc,   Mic, speaker & Device-to-Device   (D2D) communication by using multiple sensors, acceleration and microphone by   holding two devices in one hand and randomly shaking over a period of time & Shaking   together & Impersonation \\ \hline
\cite{xu2019gait} &  & 2019 & Gait & Acc & Gait-based   shared key generation system that assists two devices to generate a common   secure key by exploiting the user’s unique walking pattern & Using   rotation matrix & Eavesdropping,   adversary \\ \hline
\cite{Lin2019} & H2B & 2019 & SCG & Piezo   Vibration Sensors & Using piezo sensors to detect heartbeat signal and generate a secret key & Detecting   piezo-based IPI & Passive   eavesdropping attacks, Active presentation attacks, Active video attacks \\ \hline
\cite{Shen2018} & Shake-n-Shack & 2018 & Motion & Acc & Enabling secure data exchange between smart wearables via handshakes & Handshaking & Mimicking   attacks \\ \hline
\cite{Schurmann2018} & BANDANA & 2018 & Gait & Acc & A secure spontaneous authentication scheme that exploits correlation in   acceleration sequences from devices worn or carried together by the same   person to extract always-fresh secure secrets. & Utilize   instantaneous variations in gait sequences with respect to the mean & MITN,         Mimic Gait, Video Recording, Attach   Malicious Device \\ \hline
\cite{Sun2017} &  & 2017 & Gait & Acc & Symmetric   key generation scheme based on the timing information of gait & Inter-pulse   Intervals (IPI) of consecutive Gait &  \\ \hline
\cite{xu2017gait} & Gait-Key & 2017 & Gait & Acc & A   shared secret key generation scheme that allows two legitimate devices to   establish a common cryptographic key by exploiting users’ walking   characteristics (gait) & Using   independent component analysis (ICA) for Blind Source Separation (BSS) & Impersonation   attack,     passive eavesdropping   adversary,  active spoofing attack \\ \hline
\cite{jin2015magpairing} & MagPairing & 2016 & Motion & Magnet- ometer & Pairing   smartphones in close proximity by exploiting correlated magnetometer readings & Detecting   correlated magnetometer readings & Passive   attacks MITM attacks, replay attacks reflection attacks \\ \hline
\cite{xu2016walkie} & Walkie-Talkie & 2016 & Gait & Acc & A   shared secret key generation scheme that allows two legitimate devices to   establish a common cryptographic key by exploiting users’ walking   characteristics (gait) & Using   independent component analysis (ICA) for Blind Source Separation (BSS) & Impersonation   attack,      passive eavesdropping   adversary, active spoofing attack \\ \hline
\cite{Yang2016} & Secret   from Muscle & 2016 & EMG & EMG & A   system that can securely pair wearable devices by leveraging the electrical   activity caused by human muscle contraction, Electromyogram (EMG), to   generate a secret key & Detecting   muscle contraction & Impersonation \\ \hline
\cite{Yuzuguzel2015a} & ShakeMe & 2015 & Motion & Acc & Finding   a small set of effective features from shaking of two devices which are held   together in one hand & Shaking   together &  \\ \hline
\cite{Rostami2013} & H2H & 2013 & ECG & ECG & A   system to authenticate external medical device controllers and programmers to   Implantable Medical Devices (IMDs) & Detecting   ECG-based IPI & Active   adversaries, MITM \\ \hline
\cite{Groza2012} & SAPHE & 2012 & Motion & Acc & Shake   devices together, key exchange protocols based on accelerometer data that use   only simple hash functions combined with heuristic search trees & Shaking   together & MITM \\ \hline
\cite{LaMarca2007a} & Shake   Well Before Use & 2007 & Motion & Acc & A   method for device-to-device authentication that is based on shared movement   patterns which a user can simply generate by shaking devices together. & Shaking   together & MITM,         online attack , offline attack \\ \hline
\cite{Bichler2007a} &  & 2007 & Motion & Acc & Establish   a secure connection between two devices by shaking them together & Shaking   together &  \\ \hline
\end{tabular}
\label{tbl_signals}
\end{table*}

\section{Biometric-based pairing mechanism}
\begin{figure*}[t!]
    \centering
    \includegraphics[width=17cm,height=6cm]{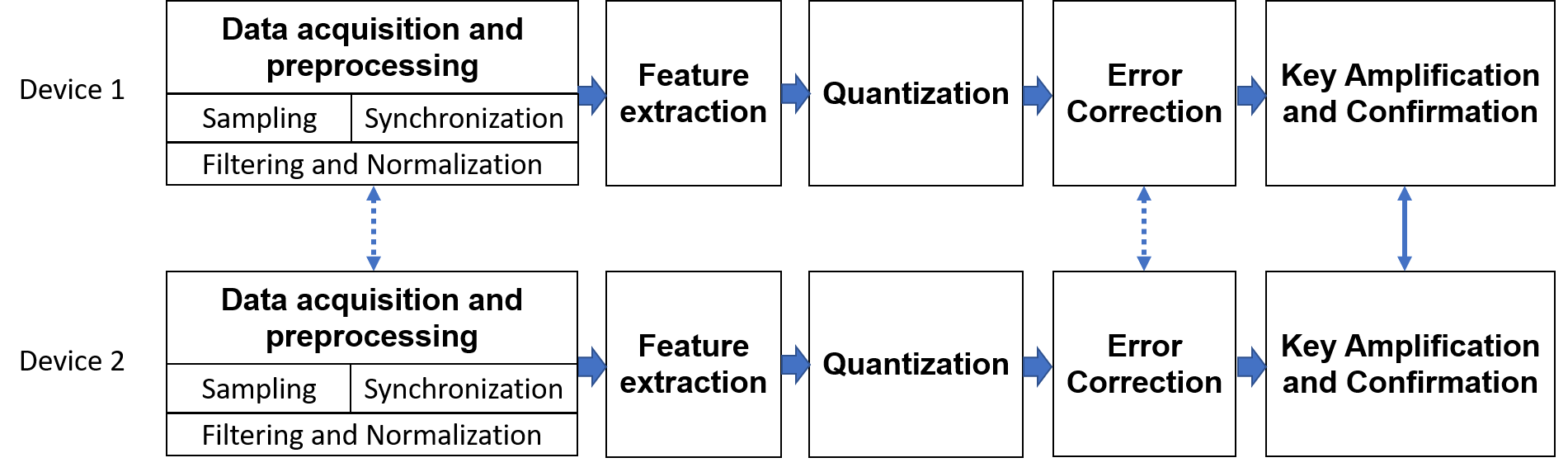}
    \caption{Context based pairing steps.}\label{steps}
 \end{figure*}

As Fig. \ref{steps} shows, sequence of signal processing is needed to separate sources (signal and noise), detect and extract features/events, quantize features/segments, correct errors, amplify bit string and create a shared key in the different devices. We explain these sequences in the following paragraphs.

\subsection{Data acquisition and preprocessing}

\subsubsection{Sampling}
Sampling is the basic stage in which raw data is collected through various sensors, from the accelerometer to the ECG sensor. The speed and quality of data acquisition can directly affect key generation. Ambient noise and system noise cause the bit current of both devices to mismatch and lead to unsuccessful pairing. On the other hand, the amount of sampling affects the amount of entropy extracted from the sensors' values. This is because the sampling rate determines the accuracy of the measurement: the higher the sampling rate, the more accurate the measured signal, the more noise \cite{Iftekharuddin2012}. Therefore, as the sampling rate increases, the measurements will have more entropy.

\subsubsection{Synchronization}
Synchronization is required to ensure all legitimate devices capture samples at the same time. Since devices are unsynchronized by default, they must agree on a common starting point. Time asynchrony would result in a high bit string mismatch rate between devices after bit quantification, which would decrease the probability of successful key distribution. Since both smartphones are unsynchronized initially, they need a user interaction for synchronizing the starting points of recording the shaking process. This task can be realized through direct user input, such as pressing a button. Alternatively, synchronization can be done by using a coordinator server \cite{Shen2018, schurmann2011secure}. Hence, upon receiving the coordinator's synchronization signal, all the sensor nodes in the same network will start recording data \cite{Sun2017}. However, we cannot ensure this kind of synchronization is always available for users. The wireless communication technique also can be exploited for synchronization purposes, e.g., a time-slotted channel hopping-based (TSCH) communication leads to temporal alignment between devices \cite{elsts2016microsecond}. Synchronization can be at a sample level, i.e., within less than half the sample width, or at an event level, i.e., based on the onset of detected (explicit or implicit) events with the respective device. Depending on the sensors used for pairing, different events can be considered as anchor points. For instance, the event can be a heel-strike \cite{xu2017gait}, bumping the devices or shaking them together in one hand \cite{Yuzuguzel2015a}.
Although better synchronization reduces the bit mismatches at the next stage, the pairing protocol should support devices without a screen and keyboard and not impose strict synchronization on pairing devices when collecting ambient context information \cite{Yang2016}.

\subsubsection{Filtering and Normalization}
The sensor data must pass a filtering step to eliminate the relative effect of ambient and artificial noise and extract the desired signal (motion, gait, PPG, ECG, EMG, etc.) from the raw data. Depending on the signal type, the filtering stage involves using a high-pass filter, a mid-pass filter, or a low-pass filter. For best results, the cut-off frequency of these filters should be in the range of the minimum and maximum frequency of the desired signal; for example, 3 Hz can be a suitable cut-off frequency for a low-pass filter for gait signal because the normal step frequency lies between 1.6-2.8 Hz \cite{xu2019gait}. In fact, by filtering the raw signal, the unwanted frequency components are removed. Since the magnitude ranges of sensors are quite different, after filtering, the signal is usually normalized to have a mean of zero and a variance of one.

\subsection{Feature extraction}
Similar raw signals lead to similar feature signals. The feature extraction process maintains the user's desired characteristics (for example, heartbeats IPI, respiration rate, gait, movement) and eliminate irrelevant noises \cite{DelRosario2015}. Feature extraction is one of the most important steps in a pairing that significantly affects the bit generation process and its performance. In particular, selecting the appropriate features plays a key role in pairing because authentication is done by comparing the bits generated based on the extracted features.
Features can be selected from both time and frequency domains \cite{Naseer2019}. Time domain features can be a wide variety of features including statistical characteristics \cite{chen2008online,he2013physical,maurer2006location} such as mean; standard deviation; variance; median; root mean square; maximum; minimum; Skewness; kurtosis; crest factor; number of peaks; Peak-to-peak amplitude; zero crossing rate \cite{saunders1996real}; signal magnitude area \cite{bouten1997triaxial}; interquartile range \cite{parkka2006activity,moncada2014activity}. In the frequency domain we can select features like energy, entropy \cite{ehatisham2017authentication}, maximum frequency index, mean frequency, fast Fourier transform coefficients \cite{huynh2005analyzing}, to name but a few.

\subsection{Quantization}
Quantization represents a source's output with a large (possibly infinite) alphabet with a small alphabet. It is a many-to-one mapping and, therefore, irreversible. Quantization can be done on a sample-by-sample basis, or it can be performed on a group of samples (by dividing the signal into segments). Quantizer encodes each sample/segment by specifying the range value at multiple levels. In this process, the raw signals or features will be quantized into bit vectors. The quantizer has a significant role in the security of the generated key. By making a biased quantization, a brute force attack would become feasible. On the other hand, the quantizer can change the length of the final key. The quantizer maps each sample to a bit string whose length depends on the quantization levels. A quantizer that has $2^m$ quantizing levels can map each sample to $m$ bits \cite{Iftekharuddin2012}.

The bit representation can also affect security. The quantizer can map samples to either binary code or Gray code \cite{sayood2017introduction}. A Gray code encodes numbers so that adjacent numbers have a single-digit differing by one. Therefore, in some cases, the Gray code can perform better in detecting noise in two consecutive samples \cite{Rostami2013}. In the recent pairing methods, various type of quantization approaches has been exploited, including pairwise nearest neighbor (PNN) quantization \cite{Bichler2007a}, standard decimal-to-binary quantizer \cite{jin2015magpairing,Yuzuguzel2015a}, decimal-to-Gray-code quantizer \cite{Rostami2013,Lin2019}, uniform quantizer \cite{Sun2017}, sigma-delta quantizer \cite{Groza2020a}, and exploiting multiple thresholds \cite{Shen2018,Bejder2020,jiang2019shake,Cao2019,sen2020vibering,Groza2012,LaMarca2007a,Schurmann2018,xu2016walkie,xu2019gait}.

\subsection{Error Correction}
Due to the measured noise, there are usually mismatches in the quantized bits between the bit vectors produced by the two devices. Therefore, in the reconciliation stage, devices exchange a certain amount of information to correct all mismatches and generate a bit-by-bit matching key. We describe some of the most important error correction approaches used in wearable device pairing in the following.
\subsubsection{Index checking}
This technique exchanges the index of the valid bit positions to reach a mutual agreement on which bits will be used in the final keys. For example, suppose the key generated by Alice’s devices is $[110xx11x00]$, while for Bob is $[1100x11xx0]$, where $x$ means the position where no valid bit is presented. Then both Alice and Bob inform each other the positions of the valid bits, i.e., Alice sends $P_{Alice} = \{1, 2, 3, 6, 7, 9, 10\}$, and Bob sends $P_{Bob} = \{1, 2, 3, 4, 6, 7, 10\}$. Upon receiving the positions, they compare the received vector with the local one and agree that only the bits that are valid according to both vectors should be used. In this example, the agreed positions should be $\{1, 2, 3, 6, 7, 10\}$ so that the final symmetric keys are $[110110]$. This error correction technique is utilized in \cite{Shen2018,xu2016walkie,xu2019gait}.

\subsubsection{Error Correction Code (ECC)}
Error correction code (ECC) is commonly used to control data errors through unreliable or noisy communication channels \cite{Ho2013}. The central idea is that the sender encodes the message with additional information in an ECC form. The redundancy allows the receiver to detect a limited number of errors that may occur anywhere in the message, and often to correct these errors without retransmission \cite{wikiECC2}. Bose–Chaudhuri–Hocquenghem (BCH) \cite{bose1960class}, Reed-Solomon (RS) \cite{reed1960polynomial}, Hamming \cite{hamming1950error}, and binary Golay Codes \cite{golay1949notes} are some of the most common ECC used in the recent pairing methods \cite{Cao2019,Schurmann2018,Sun2017,xu2017gait,sen2020vibering,Yang2016}. BCH codes form a class of cyclic error-correcting codes constructed using polynomials over a finite field (also called Galois field). One of BCH codes' key features is that there is precise control over the number of symbol errors that can be corrected by the code when code design. Specifically, it is possible to design binary BCH codes can be designed that can correct multiple bit errors \cite{wikiBCH2}. Reed-Solomon codes are the subset of BCH codes among the most powerful known classes of linear, cyclic block codes. Reed Solomon describes a systematic way of building codes that could detect and correct multiple random symbol errors. By adding $t$ check symbols to the data, RS code can detect any combination of up to $t$ erroneous symbols or correct up to $t/2$ symbols. In addition, RS codes are suitable as multiple-burst bit-error correcting codes because a sequence of $b + 1$ consecutive bit errors can affect up to two symbols of size $b$ \cite{shrivastava2013error}.
Hamming code is a block code that can detect up to two simultaneous bit errors and correcting single-bit errors.
Binary Golay code is another linear error-correcting code in which a codeword is formed by taking 12 information bits and appending 11 check bits.

\subsubsection{Fuzzy cryptography}
The fuzzy cryptographic scheme enables the compatibility of a certain amount of tolerable noise between the keys extracted from different devices by changing the error correction parameters and the length of the samples used. "Fuzzy," in this context, refers to the fact that the fixed values required for cryptography will be extracted from values close to but not identical to the original key, without compromising the security required \cite{sahai2005fuzzy,anees2018discriminative}. Jiang \emph{et al.} \cite{jiang2019shake} used this technique for error-correcting in their pairing algorithm. In \cite{Sun2017}, Sun \emph{et al.} exploited fuzzy cryptography and BCH to provide the superior performance of false acceptance rate (FAR).

\subsubsection{Compressive Sensing}
Compressed sensing theory has shown that sparse signals can be reconstructed exactly from remarkably few measurements \cite{Donoho2006}. Hence, it can be exploited as an error correction method \cite{Chartrand2007, Zoerlein2013}. 
Compressed sensing is a technique where a signal $x$ is multiplied by an $M \times N (M < N)$ sampling matrix to be sampled and compressed in a single operation. The signal $x$ is recovered by finding the $l1$ norm of the sparse version of $x$, represented by the received samples $y$. In other words, out of the infinite signals which could have been used to create the received $y$, the sparsest one is recovered as the original signal. As long as $x$ is "sparse enough," it can be recovered exactly using this technique. Lin \emph{et al.} used this method in their error correction method \cite{Lin2019}.

\subsection{Key Amplification}
In the reconciliation phase, the devices may send some information to each other through a public wireless channel. Besides, some slightly different samples/segments may have the same bit string due to error correction. Thus, an adversary can infer some private information about the secret sequence. The privacy amplification process solves this issue. Key amplification helps to increase the final key's randomness to eliminate the information leakage and increase entropy. Typically, two methods are used to combine keys generated from different segments and eliminate the correlation between them, bitwise XOR function \cite{xu2016walkie,xu2017gait,xu2019gait} and hash function MD5 and SHA-256 \cite{Bejder2020,LaMarca2007a,Bichler2007a,Cao2019,Schurmann2018,Rostami2013}.

\subsection{Performance metrics}
The ultimate objective of pairing is to generate the same key on different devices independently. On the other hand, due to wearable devices' hardware limitations, all the operations in such devices are expected to use as little memory and computational power as possible and communicate the smallest amount of data with the smallest number of messages to achieve the smallest overall energy consumption. There are several evaluation metrics for evaluating pairing system performance which we have provided a brief description of them in the lines below \cite{Hosseinzadeh2021, Abuhamad2021, Kompara2018}.

\begin{itemize}
\item \emph{Bit generation rate:}\\
The number of bits generated from the sensor readings per second.
\item \emph{Key generation rate:}\\
The major performance metric for symmetric key generation is the success rate, the probability that two keys generated by Alice and Bob can completely agree with each other (The probability of 100\% matching). In other word, it is the percentage of identical keys generated by two devices in one second.
\item \emph{Bit agreement rate:}\\
Bit Agreement Rate denotes the percentage of the matching bits of the two cryptographic keys generated by two devices
\item \emph{Computational cost:}\\
The next important performance indicator is computational cost. It is important for schemes to be as computationally efficient as possible, because sensor nodes do not pose much processing power and because more computing uses up more of the very much limited energy supply. The most common method to analyze computation cost is by measuring the amount of time it takes for the necessary operations to finish processing.
\item \emph{Energy consumption:}\\
energy consumption was measured by how much energy is spent on every bit of information produced.
\item \emph{Communication cost:}\\
Measuring of communication cost is very important, because it is the most energy consuming operation of them all. The most common ways of determining the communication cost are by the size of the sent data
\item \emph{False positive ratio:}\\
 False positive ratio (FPR), also known as false accept rate (FAR), is defined as the percentage of pairing attempts that incorrectly generate common keys among all the expected unsuccessful attempts. The metric indicates the likelihood of the adversary successfully pairing with a legitimate device by the adversary. The FPR can be computed as:$FPR=\frac{FP}{EN}$, where $FP$ is the number of incorrectly generated keys, and $EN$ is the number of expected unsuccessful attempts.
\item \emph{False negative ratio:}\\
False negative ratio (FNR), also known as false reject rate (FRR), is defined as the percentage of incorrectly unsuccessful pairing attempts among all the expected successful pairing attempts. It indicates the probability that two legitimate devices attached to the same body can not pair successfully. FNR is computed as $FNR=\frac{FN}{EP}$, where $FN$ is the number of incorrectly missed keys, and $EP$ is the number of expected successful attempts.

\item \emph{Equal/crossover error rate (EER/CER):}\\
Equal or crossover error rate (EER/CER) is the rate at which both acceptance and rejection errors are equal. The value of the EER can be easily obtained from the intersection point between the FAR and the FRR curves. Equal Error Rate (EER) measures the trade-off between FAR and FRR and it is the value of FAR or FRR when the two false rates are equal.
\item \emph{Entropy:}\\
Another important security metric in the pairing schemes is the entropy estimation. Entropy is the measure of uncertainty or randomness in the bit string generated from the measured signals. Higher entropy means more randomness to the generated bit string or in other words less dependencies between the bits. Some papers use the NIST test suite \cite{rukhin2001statistical} to estimate the entropy. 
\end{itemize}

A list of various processes and techniques used in wearable devices pairing techniques is shown the Table \ref{tbl_process}.  

\begin{table*}[]
\centering
\caption{Processes and techniques utilized in wearable device pairing studies }
\begin{tabular}{|l|p{1.9cm}|p{1.5cm}|p{1.9cm}|p{1.4cm}|p{1cm}|p{1.8cm}|p{1cm}|p{1.5cm}|p{1cm}|}
\hline
\textbf{Scheme} & \textbf{Synchronization} & \textbf{Filtering} & \textbf{Feature   Extraction} & \textbf{Quantization} & \textbf{Bit Presentation} & \textbf{Error Correction} & \textbf{Amplifi- cation} & \textbf{Confirmation} & \textbf{Encryp- tion} \\ \hline
\cite{Shen2018} & First peak &  & Dominant motion   features, PCA & Thresholds & Binary & Index exchange &  &  & AES \\ \hline
\cite{Bejder2020} & Using gateway & Low-pass filter 2 Hz,   mean filter &  & Hysteresis   thresholding & Binary & Error correction code & Hash function & Challenge-Response   (CR) mechanism &  \\ \hline
\cite{jiang2019shake} & First sample with   magnitude of 0 & Low-pass filter 5Hz & Acceleration   magnitude of 0 & Thresholds & Binary & Fuzzy cryptography &  & MAC (message   authentication code) &  \\ \hline
\cite{Groza2020a} & Exchange time message   by an unsecure channel & Low-pass filte and   High-pass & Sign of each sample & Sigma-delta & Binary &  &  &  & EKE and SPEKE \\ \hline
\cite{Cao2019} & First extreme point & Filtering Algorithm   (EPEFA) & Extreme points   extracting & Thresholds & Binary & BCH & Hash function MD5 & Verify over audio and   RF & AES \\ \hline
\cite{Yuzuguzel2015a} & Bumping both devices & Low-pass FIR filte & 10 statistics   features & Decimal-to-binary & Binary &  &  &  &  \\ \hline
\cite{sen2020vibering} &  & Band-pass filter & 5 statistics features & Thresholds & Binary & Hamming &  &  & AES \\ \hline
\cite{Groza2012} &  &  &  & Thresholds & Binary & Hashed heuristic tree &  &  &  \\ \hline
\cite{LaMarca2007a} & Onset of detected   events &  & Coherence and   quantized FFT coefficien & Thresholds & Binary &  & Hash function & DH, CKP & AES \\ \hline
\cite{Bichler2007a} & Peak of   cross-correlation function & Low-pass filte & Weight vector of   segments, PCA & Pairwise nearest   neighbor & Binary &  & Hash function &  &  \\ \hline
\cite{jin2015magpairing} & Tap two devices &  &  & Decimal to binary & Binary &  &  & Diffie-Hellma & AES \\ \hline
\cite{Sun2017} &  & Low-pass filte & Peaks of signal -   IPIs & Uniform quantizer   with q levels & Gray & Fuzzy commitment and   BCH &  &  &  \\ \hline
\cite{Schurmann2018} &  & Madgwick and   Chebyshev bandpass filte & Gait cycle detection & Threshold & Binary & BCH & Hash function & Password   Authenticated Key Exchanges (PAKE), fuzzy cryptography & SHA-256 \\ \hline
\cite{xu2016walkie} & First heel-strike   event & Low-pass filter &  & Thresholds & Binary & Index exchange & Bit-wise XOR function & MAC & AES \\ \hline
\cite{xu2017gait} & First heel-strike   event & Low-pass filter &  & M-ary quantization & Binary & Reed-Solomon (RS) & Bit-wise XOR function & MAC & AES \\ \hline
\cite{xu2019gait} & First heel-strike   event & Low-pass filter &  & Thresholds & Binary & Index exchange & Bit-wise XOR function & MAC, HMAC-MD5 & AES \\ \hline
\cite{Rostami2013} &  & Discarding four bits   of IPI & Inter-pulse interval   (IPI) & Decimal to gary code & Gray &  & Hash function & Neyman Pearson Lemma & AES-128, RSA \\ \hline
\cite{Lin2019} & Time-slotted channel   hopping-based (TSCH) & Savitzky-Golay (SG)   filter, discarding the least 3 bits of IPI & Inter-pulse interval   (IPI) & Decimal to gary code & Gray & Compressive Sensing   (CS) &  &  &  \\ \hline
\cite{Yang2016} &  & High-pass filte,   notch filter, Chebyshev IIR filter & Rectified EMG signal & Fast dynamic time   warping with three levels & Binary & Binary Golay Code &  &  &  \\ \hline
\end{tabular}
\label{tbl_process}
\end{table*}

\section{Adversary model/ Attack scenarios}
Wearable device pairing can face various attacks largely due to the broadcast nature of the wireless communication between wearable devices. We must consider the presence of a strong attacker during key generation. Eve is fully aware of the system and control of the communication channel, meaning that she may monitor, jam, and modify messages at will. Various attacks on wearable devices pairing can be categorized into passive and active attacks \cite{yang2017survey,deogirikar2017security, Narwal2021}. In this section, we provided a brief description of these attacks and discussed some common countermeasures.

\subsection{Passive attacks}
A passive attacker monitors the network for information but does not affect the target network. Passive attacks are performed as a preliminary act for the active attack \cite{kao2006eavesdropping}.
Data Sniffing/Eavesdropping: Data Sniffing or Snooping attack is an old security issue. Sniffing is an incursion that involves a weak connection between the WBAN node and the server. The attacker passively accesses the data traffic (important health data, routing updates, node ID numbers, etc.) for later analysis by sitting between the unsecured network paths. 
Detecting a passive attack is exceedingly difficult and impossible in many cases because it does not involve any changes. However, protective measures can be implemented to stop it, including:
Avoid posting sensitive information publicly, using random key distribution and strong encryption techniques to scramble messages, making them unreadable for any unintended recipients. 

\subsection{Active attacks}
An active attack involves using information gathered during a passive attack to compromise a user or network. Active attackers can cause devastation to the system as they attempt to intercept the wireless communication to change the information present on the target or en route to the target.
There are several types of active attacks. In an impersonation/spoofing attack, an attacker pretends to be another user to access the system's restricted area. In a replay attack, the intruder steals a packet from the network and forwards that packet to a service or application as if the intruder were the user who originally sent the packet. Denial-of-service (DoS) and distributed denial-of-service (DDoS) attacks are also examples of active attacks, both of which work by preventing authorized users from accessing a specific resource on a network or the internet (for example, flooding a device with more traffic than it can handle).

Unlike a passive attack, an active attack is more likely to be discovered quickly by the target upon executing it.
For instance, we can stop the attacker from impersonating the nodes by using authentication mechanisms and intrusion detection. To defend against reply attack, nonces and time token can be used to introduce fresh data \cite{malladi2002preventing}.
Authentication and anti-replay protection are the solutions suggested for avoiding denial of service \cite{wang2006survey, walters2007wireless}.

\section{Limitations and challenges}
The wearable device pairing methods should be lightweight with fast computation, low storage, and low transmission overhead. Otherwise, the power and storage space of the body sensors could be drained quickly. There are different limitations and challenges when trying to pair devices using sensors’ data. The main limitations of the current approaches are that they require the user to do a specific action (e.g., walking, handshake, gesture), and some devices should be placed on a certain part of the body (e.g., wrist, arm, head) to collect the desired signal. Also, some techniques need special sensors like ECG and EMG sensors which are not used in common wearable devices. These requirements can limit the usability of the proposed techniques. On the other hand, some other challenges can affect the accuracy of proposed methods; user's motion artifact and health condition can challenge the pairing's success rate. In some pairing methods, users must be static during the data collection phase. Furthermore, a recent study \cite{bruesch2019security} revealed gait-based pairing approaches are vulnerable to video attack.

\section{Conclusion}
By the rapid growth of smart wearable devices, pairing has become the first and the most important concern in establishing a secure communication. Increasing various type of sensors available in these devices have enabled autonomous pairing by exploiting common features of signals based on user’s activities and biometrics such as heartbeat, respiration, gait, and motion. Accordingly, the number of research articles related to wearable device pairing has been increasing exponentially.

In this article, we have surveyed recent approaches in wearable device pairing and classified them according to the signal types they have been used. We also descirebed the common required steps needed in most approaches to generate a common key in different devices. We then compared the features and tools that authors used in each step. In addition, we mentioned the adversary models and countermeasures.
Finally, we discussed challenges and limitations in pairing wearable devices. This survey is expected to be useful for the enhancement of spontaneous wearable device pairing.

\bibliographystyle{IEEEtran}
\bibliography{IEEEabrv,refs}

\begin{IEEEbiography}[{\includegraphics[width=1in,height=1.25in,clip,keepaspectratio]{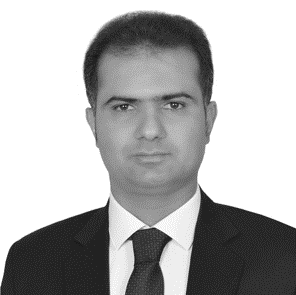}}]{JAFAR POURBEMANY} received the BSc degree in Electronics and the MSc degree in Communications 
from Yazd University, in 2009 and 2014, respectively. He is currently pursuing the Ph.D. degree in Computer Science with the 
Network Security and Privacy Research Laboratory, Cleveland State University, USA, where he is currently a Research Assistant. 
His research interests include network security, body area networks, wireless sensor networks, and machine learning. 
\end{IEEEbiography}

\begin{IEEEbiography}[{\includegraphics[width=1in,height=1.25in,clip,keepaspectratio]{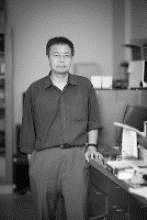}}]{YE ZHU} received the BSc degree in 1994 from Shanghai JiaoTong University 
and the MSc degree in 2002 from Texas A\& M University. He received the PhD degree from the Electrical and Computer Engineering Department at Texas A\& M University. 
He is currently an associate professor in the Department of Electrical and Computer Engineering at Cleveland State University. 
His research interests include network security, traffic engineering, and wireless sensor networks. 
He is a member of the IEEE and the IEEE Computer Society.
\end{IEEEbiography}

\begin{IEEEbiography}[{\includegraphics[width=1in,height=1.25in,clip,keepaspectratio]{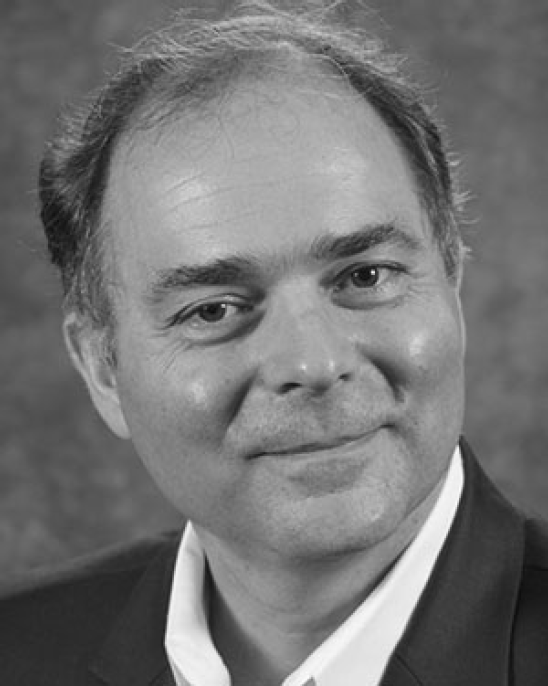}}]{Riccardo Bettati} received the diploma in informatics
    from the Swiss Federal Institute of Technology
    (ETH), Zuerich, Switzerland, in 1988 and
    the PhD degree from the University of Illinois at
    Urbana-Champaign, in 1994. He is a professor
    with the Department of Computer Science, Texas
    A\&M University, where he has been leading the
    Real-Time Systems Research Group and until
    2015 the Center for Information Assurance and
    Security. From 1993 to 1995, he held a postdoctoral
    position with the International Computer
    Science Institute in Berkeley and with the University of California at
    Berkeley. His research interests include traffic analysis and privacy,
    real-time distributed systems, real-time communication, and network
    support for resilient distributed applications. He was the program and
    general chairs of The IEEE Real-Time and Embedded Technology and
    Applications Symposia in 2002 and 2003, respectively. He shares Best
    Paper awards with collaborators and students in the IEEE National Aerospace and Electronics Conference and in the Euromicro Conference on
    Real-Time Systems.
\end{IEEEbiography}

\end{document}